\documentstyle[aps,12pt]{revtex}

\begin{document}
\title{Controllable Persistent Atom Current of Bose-Einstein Condensates in an
Optical Lattice Ring\thanks{%
The project supported by National Natural Science Foundation of China under
Grant No. 10475053.}}
\author{Zheng Gong-Ping and Liang Jiu-Qing}
\address{Institute of Theoretical Physics and Department of Physics, Shanxi\\
University, Taiyuan, Shanxi 030006, China}
\maketitle

\begin{abstract}
{\bf Abstract:} In this paper the macroscopic quantum states of
Bose-Einstein condensates in optical lattices is studied by solving the
periodic Gross-Pitaevskii equation in one-dimensional geometry. It is shown
that an exact solution seen to be a travelling wave of excited macroscopic
quantum states resultes in a persistent atom current which can be controlled
by adjusting of the barrier height of the optical periodic potential. A
critical condition to generate the travelling wave is demonstrated and we
moreover propose a practical experiment to realize the persistent atom
current in a toroidal atom waveguide.
\end{abstract}

\begin{quotation}

\end{quotation}

Dramatic features of fundamental and applied importance for Bose-Einstein
condensates (BECs) loaded in a periodic potential have attracted much
interest recently where the periodic potential is created by a laser
standing wave known as optical lattice. For example the direct observation
of an oscillating atomic current in a one-dimensional array of Josephson
junctions is realized with atomic BECs trapped in the valleys of the
periodic potential and weakly coupled by the interwell barriers.$^{[1,2]}$
The experimental realization of BECs of weakly interacting atoms has
provided a route to study atom current in a controllable and tunable
environment. The possibility of loading a BEC in a one-dimensional periodic
potential admits the observation of quantum phase effects on a macroscopic
scale such as quantum interference and the study of superfluidity on a local
scale. The superfluid array may allow investigation of phenomena so far
inaccessible to superconducting Josephson junctions and lays a bridge
between the condensate dynamics and the physics of discrete nonlinear media.

Moreover the BEC trapped in an optical lattice exhibits a novel feature that
is the quantum phase transition between Mott-insulator and superfluid.$%
^{[3]} $ As a matter of fact if the corrugated potential is shallow enough
that tunnelling establishes phase coherence across the array the atomic gas
of bosons in BEC can be kept in the superfluid phase as long as the
atom-atom interactions are small comparing with the tunnel coupling. In this
regime the kinetic energy is dominant in total energy of the boson system.
With increase of the potential depth of the optical lattice to a level that
suppresses the transport between adjacent wells the system attends an
insulator phase above a critical value of the potential depth. In this case
the phase coherence is absent and the number of boson atoms in each lattice
site becomes the same. The system then possesses a Mott-insulator behavior.
Theoretically the equation of motion for the system which realizes the array
of weakly coupled condensates satisfies a discrete nonlinear Schr\"{o}dinger
equation (NSE). Various approaches have been proposed to understand
theoretically the quantum phase transition and to determine phase diagram as
a function of BEC parameters.$^{[4-6]}$

The soliton solution of Gross-Pitaevskii(GP) equation (or NSE) describing a
continuous BEC in the mean field approximation is also studied.$^{[7-11]}$
It is well known that the periodic potential in NSE allows to stabilize
bright solitons which are not admitted for defocusing nonlinear system. The
formation of bright solitons of the NSE has been recently demonstrated for
repulsive BECs in optical lattices.$^{[7-11]}$ The soliton solutions of NSE
with periodic potential correspond to localized states of excitations with
energies inside the gaps of underlying linear band structure called gap
solitons.$^{[8]}$ On other hand a new family of stationary solutions of NSE
with periodic potential which is not in the tight-binding regime has been
constructed by Bronski, Carr, Deconink and Kutz (BCDK)$^{[12]}$ and is shown
to model a quasi-one-dimensional dilute gas BEC trapped in a standing light
wave. The BCDK solution is interesting and fruitful of physics. In this
paper we following BCDK$^{[12]}$ re-study the exact solution of the NSE for
BEC in an optical periodic potential for both repulsive and attractive
interatom collisions. It is observed that the solution is seen to be a
travelling wave and leads to a persistent atom current. Since the periodic
solution in one-dimensional space with spatial period $L$ is equivalent to
the solution in a ring of circumstance $L$ the steady current can be
realized in an optical lattice ring.

We consider a BEC confined by a cylindrically symmetric harmonic magnetic
trap with an optical periodic potential in the axial direction. The
normalized GP equation or more generally the NSE for BEC with the motion in
the radial direction essentially frozen$^{[13,14]}$ to the ground state of
the harmonic trap is seen as:$^{[15]}$ 
\begin{equation}
i\hbar \frac{\partial \psi }{\partial t}=-\frac{\hbar ^2}{2m}\frac{\partial
^2\psi }{\partial x^2}+g\left| \psi \right| ^2\psi +V(x)\psi \text{,}
\end{equation}
where the coupling constant of two-atom interaction is given by $g=\frac{%
2\hbar ^2a}{ml_0^2}$ $^{[16,17]}$ with $l_0\equiv \sqrt{\hbar /m\omega _0}$
denoting the characteristic length extension of the ground state wave
function of harmonic oscillator. $m$ is the mass of the atoms and $a$ is the 
{\it s-}wave scattering length. The confinement along the radial direction
is so tight that the trap energy with frequency $\omega _0$ along the radial
direction is much greater than the mean field interaction energy. The
optical periodic potential is written as 
\begin{equation}
V(x)=V_0\text{sin}^2\left( k_Lx\right) \text{,}
\end{equation}
with $V_0$ denoting the height of potential barrier, where $k_L=2\pi
/\lambda $ is the wave vector of the laser light and $\lambda $ is the
wavelength. The corresponding lattice period is $d=\lambda /2$. For the case
of weakly coupled condensates the condensate order parameter $\psi $ can be
decomposed as a sum of wave functions localized in each well of the periodic
potential (tight binding approximation) with the assumption relying on the
fact that the height of the interwell barrier is much higher than the
chemical potential.$^{[1,2]}$ We, however, do not restrict ourself on the
low energy case and look for the global condensate wave function of
excitation$^{[12]}$ $\psi (x,t)=\phi (x)\exp \left( -i\mu t/\hbar \right) $,
where $\mu $ is the chemical potential. Thus the spatial wave function
satisfies the stationary NSE that 
\begin{equation}
\mu \phi =-\frac{\hbar ^2}{2m}\frac{\partial ^2\phi }{\partial x^2}+g\left|
\phi \right| ^2\phi +V(x)\phi \text{.}
\end{equation}
With the general form of spatial wave function $\phi (x)$ written as$%
^{[12,18]}$ $\phi (x)=r(x)\exp \left[ i\varphi \left( x\right) \right] $ the
eq. (3) is separated as real and imaginary equations from which we obtain
the first-order differential equation for the phase $\varphi $

\begin{equation}
\varphi ^{^{\prime }}(x)=\frac \alpha {r^2(x)}\text{,}
\end{equation}
and the amplitude $r$

\begin{eqnarray}
\left( rr^{^{\prime }}\right) ^2 &=&\frac{2a}{l_0^2}r^6-\frac{2m\mu }{\hbar
^2}r^4+\beta r^2 \\
&&-\alpha ^2+\frac{2m}{\hbar ^2}r^2\int V\left( x\right) d\left( r^2\right) 
\text{,}  \nonumber
\end{eqnarray}
where parameters $\alpha $ and $\beta $ are constants of integration. An
exact solution is found as

\begin{equation}
r^2(x)=A\text{sin}^2\left( k_Lx\right) +B\text{,}
\end{equation}
with the phase determined from the equation given by 
\begin{equation}
\tan \left[ \varphi \left( x\right) \right] =\pm \sqrt{1+\frac AB}\tan
\left( k_Lx\right) \text{,}
\end{equation}
where the new coming constants of integration $A$ and $B$ along with $\alpha 
$ and $\beta $ are to be determined. The constant $B$ is obviously the mean
amplitude and acts as the dc offset for the number of the condensed atoms.$%
^{[12]}$

Substituting the amplitude formula eq. (6) into the eq. (5) the parameters $%
\alpha $, $A$ and chemical potential $\mu $ are derived as 
\begin{equation}
A=-\frac{mV_0l_0^2}{2a\hbar ^2}=-\frac{V_0}{2a\hbar \omega _0}\text{,}
\end{equation}
and $\alpha ^2=k_L^2B\!\left( B\!+\!A\right) $ with the parameter $\beta $
being eliminated. The constant $B$ can be determined from the normalization
condition that $N=\int_0^L\left| \psi \left( x,t\right) \right| ^2dx$, where 
$N$ denotes the total number of atoms in the cylindrically symmetric
magnetic trap, and the result is seen to be 
\begin{equation}
B=n-\frac A2\text{.}
\end{equation}
$L$ is the length of optical lattice and the wave function is assumed to
satisfy the periodic boundary condition such that $\psi (x)=\psi (x+L)$
which can be fulfilled if the total length is an integer number of $d$. Here 
$n$ denotes the number of atoms per unit length in the BEC of cigar-like
shape. Thus the energy spectrum is obtained as 
\begin{equation}
\mu =E_R+\!2an\hbar \omega _0+\frac{V_0}2\text{,}
\end{equation}
where $E_R=\frac{\hbar ^2k_L^2}{2m}$ is called the recoil energy of an atom
absorbing one of the lattice phonons.$^{[1]}$ The second term is the energy
due to the interaction between atoms which is proportional to the number
density of atoms. It is found that the solution of travelling waves is valid
only for a condition that the number density of atoms is greater than a
critical value i. e. 
\begin{equation}
n\geq n_c\text{,}\qquad n_c=\frac{|A|}2=\frac{V_0}{4|a|\hbar \omega _0}
\end{equation}
for both repulsive and attractive interaction between atoms. The value of
chemical potential is then confined by $\mu \geq E_R+V_0$ for the repulsive
interaction and $\mu \leq E_R$ for attractive interaction for which the
energy spectrum becomes negative when the number density of atoms is greater
than a value, $n>\frac 1{4|a|\hbar \omega _0}(V_0+2E_R)$.

The condensate atom-current can be evaluated from the usual definition, $%
j=\left( \frac \hbar m\right) 
\mathop{\rm Im}%
\left( \psi ^{*}\frac{\partial \psi }{\partial x}\right) $,$^{[19]}$ with
the exact wave function and the result is 
\begin{equation}
j=\pm \frac{\hbar k_L}m\sqrt{(n-\frac A2)(n+\frac A2)}\text{,}
\end{equation}
which is a persistent current existing when the critical condition eq. (11)
is satisfied. The atom current can be controlled by adjusting of the barrier
height of the periodic potential $V_0$. The current increases with the
decrease of the barrier height and approaches the asymptotic maximum value $%
j_{\max }=\pm \frac{\hbar nk_L}m$ for both the repulsive and attractive
interactions when the barrier height becomes vanishingly small. The current
as a function of the the barrier height is plotted in Fig. 1.

In ref. [1], a one-dimensional array of Josephson Junction was realized
experimentally with optical lattice for an atomic BEC, where the height of
interwell barriers is assumed to be much higher than the chemical potential
and therefore the condensate atoms are trapped in the potential wells with a
weak coupling between adjacent wells by quantum tunneling. While in this
paper, an opposite limit is considered that the chemical potential is higher
than the barrier height for the repulsive interaction.

The persistent atom current is a critical macroscopic quantum phenomenon and
it is certainly of importance to see whether or not the persistent current
can be realized experimentally with the recent progress made on the
confinement of atoms in the light-induced$^{[20,21]}$ and
magnetic-field-induced$^{[22,23]}$ atom waveguides.$^{[24]}$ To this end we
adopt the typical experimental data in ref. [1] for the BEC of $^{87}Rb$
atoms confined by a cylindrically symmetric harmonic magnetic trap with the
radial frequency of magnetic trap being $\omega _0=2\pi \times 92Hz$ in
order to have a quantitative evaluation. A blue detuned laser standing wave
of the wavelength $\lambda =795nm$ is superimposed on the axis of the
magnetic trap and hence the cylindrical magnetic trap is divided into an
array of traps. The scattering length of $^{87}Rb$ atoms is $a=5.8nm$.
According to ref. [1] the barrier height of the periodic potential $V_0$ can
be varied from $0$ to $5E_R$. It turns out that the critical value of atom
number density to generate the travelling wave evaluated from the right hand
site of eq. (11) is $n_c\sim 2\times 10^8$/cm for the barrier height chosen
as $V_0=E_R$. This value of atom number density is too high to be achieved
in the experiment for the Josephson junction arrays with BEC$^{[1]}$ and
therefore the persistent current is blocked in this case (i.e. with the
barrier height $V_0=E_R$). To have the critical condition eq. (11) be
satisfied the barrier height ought to be suppressed to an order such that $%
V_0\sim 0.01E_R$, thus the critical atom number density, which becomes $%
n_c\sim 2\times 10^6$/cm, is seen to be in the same order as the
experimental value$^{[1]}$ and the persistent current therefore can be
generated practically with the same experimental apparatus described in
refs. [1,2]. The persistent current may be observed in a relatively higher
potential barrier if a higher value of radial frequency $\omega _0$ of
magnetic trap can be attained (the critical number density $n_c$ is
decreased with increase of the radial frequency $\omega _0$ seen from eq.
(11)).

We hence propose an experiment to test the persistent atom current of BEC
confined in a toroidal magnetic trap with a blue detuned laser standing wave
superimposed on the axis of the toroid. The emergence and variation of the
atom current can be observed over adjusting of the barrier height.

In conclusion, we observe that the travelling wave of macroscopic quantum
states for BEC in the periodic potential results in a persistent current and
demonstrate that the atom current can be controlled only by adjusting of the
barrier height. A practical experiment to generate the matter wave in a
toroidal waveguide is proposed. The concept and proposed experimental
realization of the persistent atom current for BEC may be of fundamental and
applied importance.

Figure Caption:

{\bf Fig. 1} The persistent atom-current $j$ (in unit $\frac{\hbar nk_L}m$)
as a function of the barrier height (in unit $4|a|n\hbar \omega _0$) of the
optical lattice.


\begin{references}
\bibitem{1}  F.S. Cataliotti, S. Burger, C. Fort, P. Maddaloni, F. Minardi,
A. Trombettoni, A. Smerzi, and M. Inguscio, Science {\bf 293} (2001){\bf \ }%
843.

\bibitem{2}  S. Burger, F.S. Cataliotti, C. Fort, F. Minardi, M. Inguscio,
M.L. Chiofalo, and M.P. Tosi, Phys. Rev. Lett. {\bf 86} (2001) 4447.

\bibitem{3}  M. Greiner, O. Mandel, T. Esslinger, T.W. H$\ddot{a}$nsch, and
I. Bloch, Nature {\bf 415} (2002) 39.

\bibitem{4}  M.P.A. Fisher, P.B. Weichman, G. Grinstein, and D.S. Fisher,
Phys. Rev. {\bf B40} (1989) 546.

\bibitem{5}  D. van Oosten, P. van der Straten, and H.T.C. Stoof, Phys. Rev. 
{\bf A63} (2001) 053601.

\bibitem{6}  Jun-Jun Liang, J.-Q. Liang, and W.-M. Liu, Phys. Rev. {\bf A68}
(2003) 043605.

\bibitem{7}  S. P\"{o}tting, O. Zobay, P. Meystre, and E.M. Wright, J. Mod.
Opt. {\bf 47} (2000) 2653.

\bibitem{8}  V.V. Konotop and M. Salerno, Phys. Rev. {\bf A65} (2002) 021602.

\bibitem{9}  A. Smerzi, A. Trombetton, P.G. Kevrekidis, and A.R. Bishop,
Phys. Rev. Lett. {\bf 89} (2002) 170402.

\bibitem{10}  W.M. Liu, B. Wu, and Q. Niu, Phys. Rev. Lett. {\bf 84} (2000)
2294; W.M. Liu, W.B. Fan, W.M. Zheng, J.Q. Liang, and S.T. Chui, Phys. Rev.
Lett. {\bf 88} (2002) 170408.

\bibitem{11}  W.D. Li, X.J. Zhou, Y.Q. Wang, J.Q. Liang, and W.M. Liu, Phys.
Rev. {\bf A64} (2001) 015602; Z.W. Xie, W. Zhang, S.T. Chui, and W.M. Liu,
Phys. Rev. {\bf A69} (2004) 053609.

\bibitem{12}  J.C. Bronski, L.D. Carr, B. Deconinck, and J.N. Kutz, Phys.
Rev. Lett. {\bf 86} (2001) 1402.

\bibitem{13}  A. G$\ddot{o}$rlitz, J.M. Vogels, A.E. Leanhardt, C. Raman,
T.L. Gustavson, J.R. Abo-Shaeer, A.P.

Chikatur, S. Gupta, S. Inouye, T.P. Rosenband, D.E. Pritchard, and W.
Ketterle, Phys. Rev. Lett. {\bf 87} (2001) 130402.

\bibitem{14}  M. Greiner, I. Bloch, O. Mandel, T.W. H$\ddot{a}$nsch, and T.
Esslinger, Phys. Rev. Lett. {\bf 87 }(2001) 160405.

\bibitem{15}  F. Dalfovo, S. Giorgini, L.P. Pitaevskii, and S. Stringari,
Rev. Mod. Phys. {\bf 71} (1999) 463.

\bibitem{16}  M. Olshanii, Phys. Rev. Lett. {\bf 81} (1998) 938.

\bibitem{17}  D.S. Petrov, G.V. Shlyapuikov, and J.T.M. Walraven, Phys. Rev.
Lett. {\bf 85} (2000) 3745.

\bibitem{18}  L.D. Carr, C.W. Clark, and W.P. Reinhardt, Phys. Rev. {\bf A62}
(2000) 063610.

\bibitem{19}  D.I. Choi and Q. Niu, Phys. Rev. Lett. {\bf 82} (1999) 2022.

\bibitem{20}  M.A. Ol'shanii, Y.B. Ovchinnikov, and V.S. Letokhov, Opt.
Commun. {\bf 98} (1993) 77.

\bibitem{21}  M.J. Renn, D. Montgomery, O. Vdovin, D.Z. Anderson, C.E.
Wieman, and E.A. Cornell, Phys. Rev. Lett. {\bf 75} (1995) 3253.

\bibitem{22}  J. Denschlag, D. Cassettari, and J. Schmiedmayer, Phys. Rev.
Lett. {\bf 82} (1999) 2014.

\bibitem{23}  A.E. Leanhardt, A.P. Chikkatur, D. Kielpinski, Y. Shin, T.L.
Gustavson, W. Ketterle, and D.E. Pritchard, Phys. Rev. Lett. {\bf 89} (2002)
040401.

\bibitem{24}  M.A. Kasevich, Science {\bf 298} (2002) 1363.
\end{references}
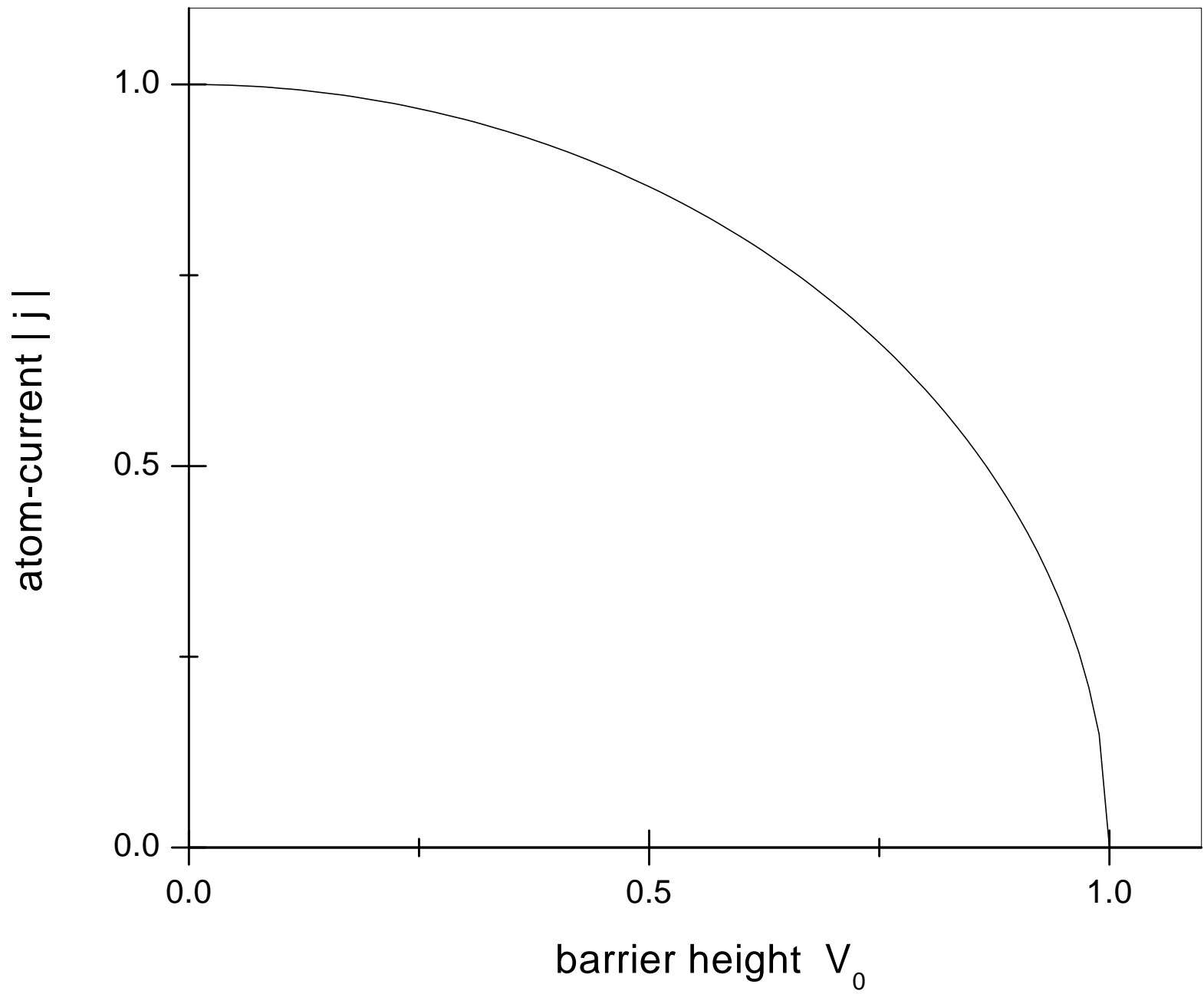
\end{document}